\documentclass{aipproc}
\layoutstyle{6x9}

\newcommand{\be}{\begin{eqnarray}}
\newcommand{\ee}{\end{eqnarray}}

\begin{document}

\vspace*{-0.5cm}
\begin{flushright}
OSU-HEP-09-07\\
\end{flushright}
\vspace{0.5cm}


\title[]{Fermion Mass Hierarchy and\\New Physics at the TeV Scale}
\author{S. Nandi\footnote{Plenary talk presented at the 17th International Conference on Supersymmetry and the Unification of
Fundamental Interactions (SUSY09) at Northeastern University,
Boston, MA, 5-10 June, 2009.}}{address={Department of Physics and
Oklahoma Center for High Energy Physics,\\Oklahoma State University,
Stillwater, OK 74078},email={s.nandi@okstate.edu}}



\begin{abstract}
In this talk, I present a new framework to understand the
long-standing fermion mass hierarchy puzzle. We extend the Standard
Model gauge symmetry by an extra local $U(1)_S$ symmetry, broken
spontaneously at the electroweak scale. All the SM particles are
singlet with respect to this $U(1)_S$. We also introduce additional
flavor symmetries, $U(1)_F$'s, with flavon scalars $F_i$, as well as
vectorlike quarks and leptons at the TeV scale. The flavon scalars
have VEV in the TeV scale. Only the top quark has the usual
dimension four Yukawa coupling. EW symmetry breaking to all other
quarks and leptons are propagated through the messenger field, $S$
through their interactions involving the heavy vector-like fermions
and $S$, as well as through their interactions involving the
vector-like fermions and $F_i$. In addition the explaining the
hierarchy of the charged fermion masses and mixings, the model has
several interesting predictions for Higgs decays,  flavor changing
neutral current processes in the top  and the b quark decays, decays
of the new singlet scalars to the new $Z'$ boson, as well as
productions of the new vectorlike quarks. These predictions can be
tested at the LHC.
\end{abstract}
\maketitle

\section{Introduction}

Fermion mass hierarchy is a long-standing problem in particle
physics. The charged fermion masses vary by five orders of
magnitude, while the quark mixing angles vary by almost two orders
of magnitude. There are two main approaches to understand this
puzzle
\cite{Froggatt:1978nt,attempts,Nandi:2008zw,Dobrescu:2008sz,Balakrishna:1987qd}.
This hierarchy is caused by physics at the high scale (GUT scale or
the Planck scale: so called Froggatt-Nielsen type mechanism; or this
is caused by some new physics at the TeV scale. In this work, I
present a new model for this second approach with new physics at the
TeV scale \cite{lmn}.


What are the new physics possibilities at the TeV scale?
Supersymmetry is highly motivated, and predicts new superpartners
and the Higgs boson at the TeV scale. Extra dimensions are somewhat
motivated, and predicts new Kaluza-Klein excitations at the TeV
scale. Extra $U(1)$ is somewhat string theory motivated, and
predicts new gauge boson at the TeV scale. However these are all
theory motivated. The experimental clues so far are that the charged
fermion masses are highly hierarchial, quark mixing angles are
hierarchial, and FCNC processes are strongly suppressed. The natural
question is what sort of new physics can explain these, and can be
observed at the LHC. In this talk, I present one such possibility.

In the SM, the Yukawa interactions of the fermions, in the mass
basis, are parameterized by

\begin{equation}
 L = y_{d_i} \bar{q_{iL}} d_{iR} H  + y_{u_i} \bar{q_{iL}} u_{iR}
\tilde{H} + h.c.
 \end{equation}

 with $m_{q_i} = y_{q_i} v$, where $ v = <H> $. Note that $y_t \sim 1$, whereas the Yakawa
 couplings of all the light quarks and leptons, $y_b, y_c, y_s, y_d,
 y_u, y_e, y_{\mu}, y_{\tau}$  are $\ll1$. Thus the top quark is directly
 connected to the EW symmetry breaking
sector, and has dimension 4 Yukawa interactions. The lighter quarks
are probably not directly connected to the EW symmetry breaking
sector, They may be connected via some messenger fields.

We know that the FCNC interactions among the quarks are highly
suppressed. This hints at the existence of some sort of flavor
symmetry. Furthermore, if we let all the SM fermions except
$q_{3L}$, $u_{3R}$ and H carry nonzero flavor charges, then the
dimension 4 Yukawa couplings for the light quarks with H will be
prevented. These flavor symmetries need to be global, and have to be
spontaneously broken at the TeV scale. What sort of fields in
addition to the SM we need to achieve this scenario? We shall see
that one possibility is to have vectorlike quarks, $Q$ and leptons at
the TeV scale, and additional flavor symmetries, $U_i(1)_{F_i}$.  Again,
since the light quarks and leptons are not directly connected to the
EW symmetry breaking scale, we need a messenger field to achieve
this, and a SM singlet complex scalar field with an extra $U(1)_S$
local symmetry will serve our purpose. Thus the new physics in our
scenario will involve $Q$, $S$ and $Z'$.

\section{Model and Formalism}

We extend the gauge symmetry of the SM model by a $U(1)_S$ local
symmetry and $U_i(1)_{F_i}$ global symmetries. All of the SM
fermions are neutral with respect to $U(1)_S$, while all of the SM
fermions apart from the third generation quark doublet $q_{3L}$ and
right-handed top $u_{3R}$ are charged under the global
$U_i(1)_{F_i}$. We introduce a complex scalar field $S$ which has
charge 1 under $U(1)_S$, is neutral under $U_i(1)_{F_i}$, and is a
SM singlet. We also introduce  complex scalar fields $F_i$, the
``flavons'', which have charges under $U_i(1)_{F_i}$, are neutral
under $U(1)_S$, and are  SM singlets. The Higgs field $H$ is taken
as neutral under $U(1)_S$ and $U_i(1)_{F_i}$. We assume that the
$U_i(1)_{F_i}$ charges of the SM fermions are such that only the top
quark has an allowed dimension 4 Yukawa interaction.

The $S$ field acquires a vev at the EW scale that spontaneously
breaks the $U(1)_S$ symmetry. The pseudoscalar component of $S$ is
eaten to give mass to the  $U(1)_S$ gauge boson $Z'$.  The field $S$ acts as a
messenger to both the flavor symmetry breaking and EW symmetry
breaking. The $U_i(1)_{F_i}$ symmetries are broken by the vev of the flavon scalar
fields, $F_i$ at the TeV scale. There are additional vectorlike fermions
at the TeV scale charged under both $U(1)_S$ and $U_i(1)_{F_i}$.

In this framework, the Yukawa interactions of the light fermions,
after integrating out the heavy vectorlike fermions appear as
higher dimensional operators in a hierarchial pattern given by
$$\left(\frac{S^\dagger S}{ M^2}\right)^n  \left(\frac{F_i}{M}\right)^{n_1}
\left(\frac{F_j^{\dag}}{M}\right)^{n_2} f_{ij} \bar{q_{iL}} d_{iR} H,$$ with
similar expressions for the up sector. The observed fermion mass
hierarchy and mixings are reproduced in powers of $\epsilon$
$$\epsilon = \frac{<s>}{M} \sim \frac{1}{7},$$
which we call the "little" hierarchy. We can absorb the $F/M$
dependence into field-dependent dimensionless complex couplings
$h_{ij}$, where $i$, $j$ are generation labels. The values of these
couplings we will then take to be of order 1.

In the model we propose, the observed fermions mass hierarchy is
generated from the following low energy effective interactions:
\begin{eqnarray}
{\cal L}^{\rm Yuk} &=& h_{33}^u \overline{q}_{3L} u_{3R} \tilde{H} +
\left({S^\dagger S \over M^2}\right) \left(h_{33}^d
\overline{q}_{3L} d_{3R} H + h_{22}^u \overline{q}_{2L} u_{2R}
\tilde{H}+h_{23}^u \overline{q}_{2L} u_{3R} \tilde{H}\right. \nonumber \\
&&+\left.h_{32}^u \overline{q}_{3L} u_{2R} \tilde{H}\right) +
\left({S^\dagger S \over M^2}\right)^2
\left(h_{22}^d\overline{q}_{2L} d_{2R} H + h_{23}^d
\overline{q}_{2L} d_{3R} H + h_{32}^d \overline{q}_{3L} d_{2R}
H\right.
\nonumber \\
&&\left.+h_{12}^u\overline{q}_{1L}u_{2R}\tilde{H}
+ h_{21}^u\overline{q}_{2L}u_{1R} \tilde{H} + h_{13}^u\overline{q}_{1L} u_{3R} \tilde{H}
+ h_{31}^u\overline{q}_{3L} u_{1R} \tilde{H} \right) \nonumber \\
&&+ \left({S^\dagger S\over M^2}\right)^3 \left(h_{11}^u
\overline{q}_{1L} u_{1R} \tilde{H} + h_{11}^d\overline{q}_{1L}
d_{1R} H  \right. + \left. h_{12}^d\overline{q}_{1L} d_{2R} H +
h_{21}^d
\overline{q}_{2L} d_{1R} H \nonumber\right.\\
&&+\left. h_{13}^d \overline{q}_{1L} d_{3R} H + h_{31}^d
\overline{q}_{3L} d_{1R} H \right ) + h.c. \label{ONE}
\end{eqnarray}
where all the couplings $h_{ij}$ are assumed to be of order 1.

Note that the above interactions are very similar to those proposed
in reference \cite{bn,gl}, except our interactions involve suppression
by powers of $\left({S^\dagger S \over M^2}\right)$, instead of
$\left({H^\dagger H \over M^2}\right)$.


\subsection{Fit to Fermion Masses and CKM Mixing}

The gauge symmetry of our model is the usual SM symmetry, plus an
additional $U(1)_S$ symmetry. The SM symmetry is broken
spontaneously by the usual Higgs doublet, $H$ at the EW scale. We
assume that the extra $U(1)_S$ symmetry is also broken spontaneously
at the EW scale by a SM singlet complex scalar field, $S$. The
pseudoscalar part of the complex scalar field, $S$ is absorbed by
the $Z'$ to get its mass. Thus after symmetry breaking, the
remaining  scalar fields are $h$ and $s$. Parameterizing  the Higgs
doublet and singlet in the unitary gauge as
\begin{equation}
H = \left(\matrix{0 \cr \frac{h^0}{\sqrt{2}}+v}\right)~~S = \left(\frac{s^0}{\sqrt{2}}+v_s\right),
\label{THREE}
\end{equation}
with $v \simeq 174$ GeV, and defining an additional small parameter
\begin{equation}
\beta \equiv { v \over M},
\label{FOUR}
\end{equation}
we obtain, from Eqs. (\ref{ONE}-\ref{FOUR}) the following mass matrices for the up and down quark sector:
\begin{eqnarray}
M_u = \left(\matrix{h_{11}^u \epsilon^6 & h_{12}^u \epsilon^4 &
h_{13}^u \epsilon^4 \cr h_{21}^u\epsilon^4 & h_{22}^u \epsilon^2 &
h_{23}^u \epsilon^2 \cr h_{31}^u \epsilon^4 & h_{32}^u \epsilon^2 &
h_{33}^u}\right)v, ~~~~~ M_d = \left(\matrix{h_{11}^d \epsilon^6 &
h_{12}^d \epsilon^6 & h_{13}^d \epsilon^6 \cr h_{21}^d\epsilon^6 &
h_{22}^d \epsilon^4 & h_{23}^d \epsilon^4 \cr h_{31}^d \epsilon^6 &
h_{32}^d \epsilon^4 & h_{33}^d \epsilon^2}\right)v~.
\label{FIVE}
\end{eqnarray}
The charged lepton mass matrix is obtained from $M_d$ by replacing
the couplings $h_{ij}$ appropriately. Note that these mass matrices
are the same as in Ref. \cite{bn}, and as was shown there, good fits to the
quark and charged lepton masses, as well as the CKM mixing angles are
obtained by choosing $\epsilon\sim 0.15$, and all the couplings
$h_{ij}$ of order one. To leading order in $\epsilon$, the
fermion masses are given by
\be
(m_t,\;m_c\;,m_u) &\simeq& (\vert h_{33}^u\vert,\; \vert h_{22}^u\vert\epsilon^2,\;
\vert h_{11}^u - h_{12}^uh_{21}^u/h_{22}^u\vert \epsilon^6)\,v\;,\nonumber\\
(m_b,\; m_s,\; m_d) &\simeq& (\vert h_{33}^d\vert\epsilon^2,\; \vert
h_{22}^d\vert\epsilon^4,\;
\vert h_{11}^d\vert\epsilon^6)\,v\;,\\
(m_{\tau},\;m_{\mu},\;m_e) &\simeq& (\vert h_{33}^\ell\vert\epsilon^2,\;
\vert h_{22}^\ell\vert\epsilon^4,\; \vert h_{11}^\ell\vert\epsilon^6)\,v \;\nonumber ,
\ee
while the quark mixing angles are
\be
\vert V_{us}\vert &\simeq& \left\vert \frac{h_{12}^d}{h_{22}^d} - \frac{h_{12}^u}{h_{22}^u} \right\vert\epsilon^2 \; ,\nonumber\\
\vert V_{cb}\vert &\simeq& \left\vert  \frac{h_{23}^d}{h_{33}^d} - \frac{h_{23}^u}{h_{33}^u} \right\vert\epsilon^2 \; , \\
\vert V_{ub}\vert &\simeq& \left\vert \frac{h_{13}^d}{h_{33}^d} - \frac{h_{12}^uh_{23}^d}{h_{22}^uh_{33}^d}
- \frac{h_{13}^u}{h_{33}^u} \right\vert\epsilon^4 \;\nonumber .
\ee


\subsection{Yukawa Interactions and FCNC}

Our model has flavor changing neutral current interactions in the
Yukawa sector. Using Eqs.(1-4), the Yukawa interaction matrices
$Y^{h}_u$, $Y^{h}_d$, $Y^{s}_u$, $Y^{s}_d $ for the up and down
sector, for $h^0$ and $s^0$ fields are obtained to be

\begin{eqnarray}
\sqrt{2} Y^{h}_u = \left(\matrix{h_{11}^u \epsilon^6 & h_{12}^u
\epsilon^4 & h_{13}^u \epsilon^4 \cr h_{21}^u\epsilon^4 & h_{22}^u
\epsilon^2 & h_{23}^u \epsilon^2 \cr h_{31}^u \epsilon^4 & h_{32}^u
\epsilon^2 & h_{33}^u}\right), ~~~~~ \sqrt{2} Y^{h}_d =
\left(\matrix{h_{11}^d \epsilon^6 & h_{12}^d \epsilon^6 & h_{13}^d
\epsilon^6 \cr h_{21}^d\epsilon^6 & h_{22}^d \epsilon^4 & h_{23}^d
\epsilon^4 \cr h_{31}^d \epsilon^6 & h_{32}^d \epsilon^4 & h_{33}^d
\epsilon^2}\right), \label{SIX}
\end{eqnarray}
with the charged lepton Yukawa coupling matrix $Y_\ell$ obtained
from $Y_d$ by replacing $h_{ij}^d \rightarrow h_{ij}^\ell$.

\begin{eqnarray}
\sqrt{2} Y^{s}_u = \left(\matrix{6h_{11}^u \epsilon^5\beta &
4h_{12}^u \epsilon^3\beta & 4h_{13}^u \epsilon^3\beta \cr
4h_{21}^u\epsilon^3\beta & 2h_{22}^u \epsilon\beta & 2h_{23}^u
\epsilon\beta \cr 4h_{31}^u \epsilon^3\beta & 2h_{32}^u
\epsilon\beta & 0}\right),\label{SEVENa}\\ \sqrt{2} Y^{s}_d =
\left(\matrix{6h_{11}^d \epsilon^5\beta & 6h_{12}^d \epsilon^5\beta
& 6h_{13}^d \epsilon^5\beta \cr 6h_{21}^d\epsilon^5\beta & 4h_{22}^d
\epsilon^3\beta & 4h_{23}^d \epsilon^3\beta \cr 6h_{31}^d
\epsilon^5\beta & 4h_{32}^d \epsilon^3\beta & 2h_{33}^d
\epsilon\beta}\right), \label{SEVENb}
\end{eqnarray}
with the charged lepton Yukawa coupling matrix $Y_\ell$ obtained
from $Y_d$ by replacing $h_{ij}^d \rightarrow h_{ij}^\ell$.

There are several important features that distinguish our model from
the proposal of Refs. \cite{bn,gl,Dorsner:2002wi}. i) Note, from Eqs.(\ref{FIVE}) and (\ref{SIX}), in our model, the
Yukawa couplings of $h$ to the SM fermions are exactly the same as in
the SM. This is because the fermion mass hierarchy in our model is
arising from $\left({S^\dagger S \over M^2}\right)$. This is a
distinguishing feature of our model from that proposed in
\cite{bn,gl} where the Yukawa couplings of $h$ are flavor dependent,
because the hierarchy there arises from $\left({H^\dagger H \over
M^2}\right)$. ii) In our model, we have an additional singlet Higgs boson whose
coupling to the SM fermions are flavor dependent as given in Eq.
(\ref{SEVENa}, \ref{SEVENb}). Again, this is because the hierarchy in our model
arises from $\left({S^\dagger S \over M^2}\right)$. In particular,
$s^0$ does not couple to the top quark, and its dominant fermionic
coupling is to the bottom quark. This will have interesting
phenomenological implications for the Higgs searches at the LHC.
iii) We note from Eq. (\ref{FIVE}-\ref{SIX}) that the mass matrices
and the corresponding Yukawa coupling matrices for $h$ are proportional
as in the SM. Thus there are no flavor changing Yukawa interactions
mediated by $h$. However, this is not true for the Yukawa interactions
of the singlet Higgs as can be seen from Eqs. (\ref{FIVE}) and
(\ref{SEVENa}, \ref{SEVENb}). Thus $s$ exchange will lead to flavor violation in the
neutral Higgs interactions.

\subsection{Higgs Sector and Extra $Z'$}

The Higgs potential of our model, consistent with the SM and the
extra $U(1)_S$ symmetry, can be written as
\begin{eqnarray}
 V(H,S) = -\mu^{2}_H (H^{\dag} H) - \mu^{2}_S (S^{\dag} S)
 + \lambda_H (H^{\dag}H)^2 + \lambda_S (S^{\dag} S)^2
 +\lambda_{HS}(H^{\dag}H)(S^{\dag} S).
\label{EIGHT}
\end{eqnarray}

Note that after absorbing the three components of $H$ in $W^{\pm}$ and
$Z$, and the pseudoscalar component of $S$ in $Z'$, we are left with
only two scalar Higgs, $h^0$ and $s^0$. The squared mass matrix in
the $(h^0, s^0)$ basis is given by
\begin{equation}
{\cal M}^2 = 2 v^2\left(\matrix{
    2\lambda_H          &  \lambda_{HS} \alpha \cr
     \lambda_{HS} \alpha    & 2\lambda_S \alpha^2  \cr
        }\right),
\label{NINE}
\end{equation}
where $\alpha=v_s/v$.

The mass eigenstates $h$ and $s$ can be written as
\begin{eqnarray}
 h^0 &=& h \cos\theta + s \sin\theta, \nonumber \\
 s^0 &=& - h \sin\theta + s \cos\theta,
\label{TEN}
\end{eqnarray}
 where $\theta$ is the  mixing angle in the Higgs sector.

 In the Yukawa interactions discussed above, as well as in the gauge
 interactions involving the Higgs fields, the fields appearing are
 $h^0$ and $s^0$, and these can be expressed in terms of $h$ and $s$
 using Eq. (\ref{TEN}).

 The mass of the $Z'$ gauge boson is given by
 \begin{equation}
 m^{2}_{Z'} = 2 g^{2}_E  v^{2}_s
 \label{ELEVEN}
 \end{equation}

Note that the $Z'$ does not couple to any SM particles directly. Its
coupling with the neutral scalar Higgs $h$ $(Z' h h$ coupling) is also
zero. The $Z'$ coupling to the SM particles will be only via dimension
six or higher operators. Such couplings are generated by the
vectorlike fermions in the model.

\section {Phenomenological Implications: Constraints, predictions
and new physics signals}

\textbf{Constraint on the mass of s:} Experiments at LEP2 have set a
lower limit of $114.4$ GeV for the mass of the SM Higgs boson. This
is due to the nonobservation of the Higgs signal from the associated
production $e^+ e^- \rightarrow Zh$. In our model, since the singlet
Higgs can mix with the doublet $h$, there will be a limit for $m_s$
depending on the value of the mixing angle, $\theta$. For
$\cos^2\theta\ge 0.25$, the bound of $114.4$ applies also for $m_s$
\cite{lpew}. However, $s$ can be lighter if the mixing is small.

\textbf{Constraint on the mass of the $Z'$:} We have assumed that
the extra $U(1)$ symmetry in our model is spontaneously broken at
the EW scale. But the corresponding gauge coupling, $g_E$ is
arbitrary and hence the mass of $Z'$ is not determined in our model.
However, very accurately measured $Z$ properties at LEP1 put a
constraint on the $Z-Z'$ mixing to be $\sim 10^{-3}$ or smaller
\cite{pdg,Langacker:2008yv}. In our model, the $Z'$ does not couple to any SM
particle directly. $Z-Z'$ mixing can take place at the one loop
level with the new vectorlike fermions in the loop. The mixing angle
is

\begin{equation}
    \theta_{ZZ'}\sim \frac{g_Z g_E}{16 \pi^2} \left(\frac{m_Z}{M}\right)^2,
\label{TWELVE}
\end{equation}
where M is the mass of the vectorlike fermions with masses in the
TeV scale. Even with $g_E \sim 1$, we get $\theta_{ZZ'} \sim 10^{-4}$ or
less. Thus there is no significant bound for the mass of this $Z'$
from the LEP1 \cite{dob}.






 \textbf{Higgs signals:} As can be seen from Eq. (\ref{SIX}),
 the  couplings of the doublet Higgs $h$ to the SM fermions
 are identical to that in the SM, whereas the couplings of the
 singlet Higgs are flavor dependent. In particular, the singlet
 Higgs $s$ does not couple to the top quark, whereas its coupling to
 $(b,\tau; c,s,\mu; u,d,e)$ involve the flavor dependent factors
 $2,2; 2,4,4; 6,6,6)$ respectively. This is, of course, in the limit
 of zero mixing between $h$ and $s$. Including the mixing, these factors
 will be modified by the appropriate mixing factors. Thus our model
 will be distinguished from the SM by the fact that the Higgs couplings,
 in general, will be fermion flavor dependent.

\begin{figure}
        \includegraphics[width=0.75\textwidth,height=3in]{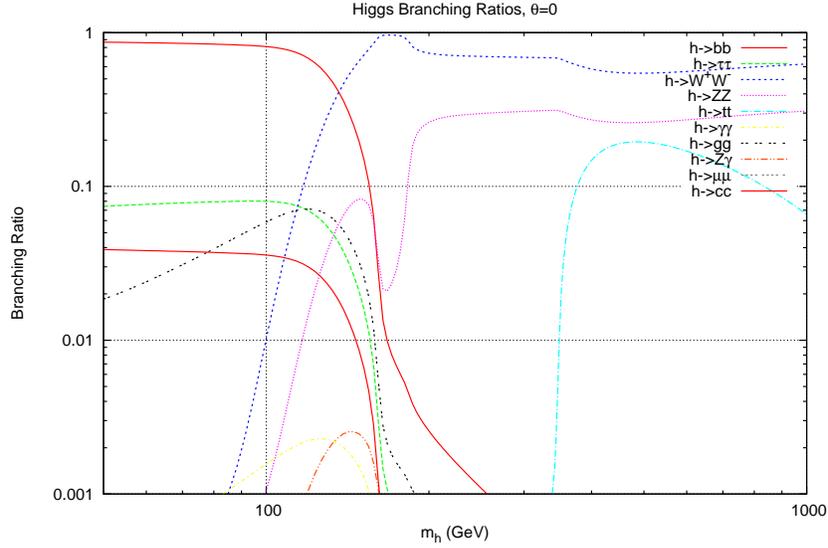}
        \caption{Branching ratio of $h\rightarrow2x$.  Here $\alpha=1$.}
        \label{h-2x_theta_0}
\end{figure}

\begin{figure}
    \includegraphics[width=0.75\textwidth,height=3in]{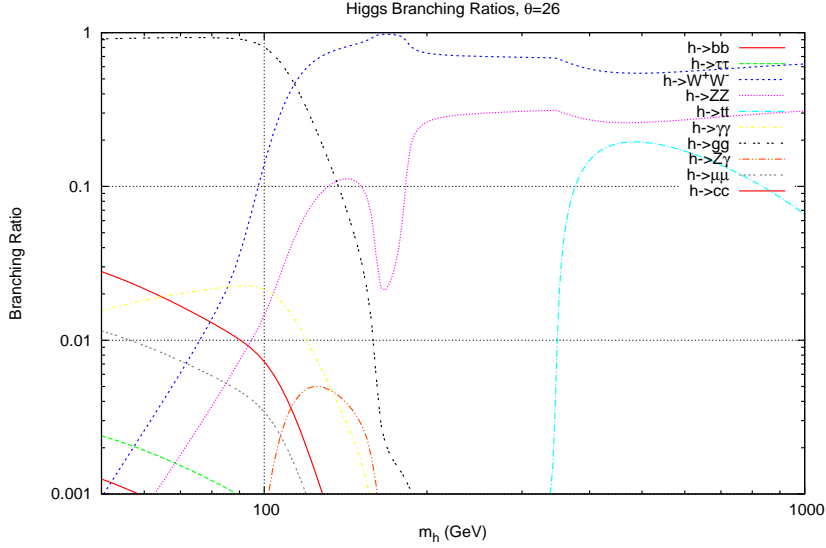}
        \caption{Branching ratio of $h\rightarrow2x$.  Here $\alpha=1$.}
        \label{h-2x_theta_26}
\end{figure}

\textbf{Higgs decays:} The couplings of the Higgs bosons $h$ and
 $s$ to the fermions and the gauge bosons can be obtained from
 Eqns. (\ref{SIX}) and (\ref{SEVENa}, \ref{SEVENb}). Because of the flavor
dependency of the couplings of $s_0$ (and hence for both $h$ and $s$
via mixing) to the fermions, the branching ratios (BR)
 for $h$ to various final states are altered substantially from those in
 the SM. These branching ratios (BR) for $h$ to  the various final
 states are shown in Figs. \ref{h-2x_theta_0}, \ref{h-2x_theta_26} for the values of
 the mixing angle, $\theta = 0^\circ$ and $26^\circ$ \cite{hdecay}.


  For $\theta = 0$, these BR's are the same as for the SM. Note
 that for $\theta =26^\circ$, the $gg$ and $\gamma \gamma$
the BR's are enhanced
 substantially compared to the SM.
 This is due to drastic reduction for the $b \overline{b}$ mode due the
 almost cancelation in the corresponding coupling
 In particular, for $\theta = 26^\circ$, the effect is quite
 dramatic. For a light Higgs ($m_h$ around $115$ GeV), the usually
 dominant $b \overline{b}$ mode is highly suppressed and the $\gamma\gamma$
 mode is enhanced by a factor of almost $10$ compared to the SM. This is
 to be contrasted with the proposal of Refs. \cite{bn,gl} in which the $h
 \rightarrow \gamma\gamma$ mode is reduced by about a factor of $10$.
 Thus the Higgs signal in this mode for a Higgs mass of $\sim{114 -
 140}$ GeV gets a big enhancement making its potential
 discovery via this mode much more favorable at the LHC. Such a
 signal may be observable at the Tevatron for a Higgs mass
 $\sim{114}$ GeV as the luminosity accumulates, but would require
 about 10 $\textrm{fb}^{-1}$ of data \cite{tevatron_search}.

 Another interesting effect is the Higgs signal via the $WW^*$  for
 the light higgs. In the SM, this mode becomes important for the
 Tevatron search staring at $m_h \sim{135}$ GeV, where the BR to
 $WW^*$ is approximate equal to that of $b\overline{b}$. Currently
 Tevatron Run2 experiments have excluded SM Higgs mass in the range of $160$ to $170$
 GeV (where the BR to $WW^*$ is around $100$ percent) for this mode.
 In our model, for $\theta = 26^\circ$ for example, note that this cross
 over between the $WW^*$ mode and the $b\overline{b}$ mode takes place
 sooner than $135$ GeV. Thus the Tevatron will be more sensitive to
 lower mass range than in the SM, and will be able to exclude mass
 ranges much smaller than $160$ GeV.

\textbf{Top quark physics: }In the SM, $t\rightarrow{c h}$ mode
 is severely suppressed with a BR $\sim10^{-14}$ \cite{tchsm}. In our model, as
 can be seen from Eqs.(\ref{SIX}) and (\ref{SEVENa}, \ref{SEVENb}), although $t\rightarrow{c h}$ is
 zero at tree level, we have a large coupling for
 $t\rightarrow{c s} \sim {2 \epsilon \beta}$. This gives rise to
 significant BR  for the $t\rightarrow{c s}$ mode for a Higgs mass
 of up to about $150$ GeV. If the mixing between the $h$ and $s$  is
 substantial, both decay modes, $t\rightarrow{c s}$ and
 $t\rightarrow{c h}$ will have BR $\sim{10^{-3}}$. With a very large
 $t \overline{t}$ cross section , $\sigma_{t\bar{t}}\sim{10^3}$ pb at the
 LHC, this can be a major discovery mode for higgs bosons at the
 LHC. Observation of signals for two different Higgs masses will also show
 clear evidence for new physics beyond the SM.

 \textbf{$\mathbf{Z^\prime}$physics:} Our model has a $Z'$ boson in
 the EW scale from the spontaneous breaking of the extra $U(1)_S$ symmetry.
 As discussed before, since the $Z-Z'$ mixing is very small
 $\sim{10^{-4}}$ or less, its mass is not constrained by the
 very accurately measured Z properties at LEP. Its mass can be as
 low as few GeV from the existing constraints. This $Z'$ does not
 couple to the SM particles with dimension 4 operators. It does
 couple to $s$ at tree level via the $sZ' Z'$ interaction. Thus it
 can be produced via the decay of $s$ (or $h$ if there is a substantial
 mixing between $h$ and $s$). This gives an interesting signal for the
 Higgs decays, $s\rightarrow{Z' Z'}$, $h\rightarrow{Z' Z'}$ if
 allowed kinematically.



 \textbf{$\mathbf{B_s^0\rightarrow\mu^+\mu^-}$:}
 In our model this decay gets a contribution from an FCNC interaction mediated
 by $s$-exchange.  The amplitude for this decay is
  $A \sim 4 h_{22}^d h_{22}^\ell \epsilon^6 \beta^2$. Taking
   $\beta\sim\epsilon$, $A \sim 4 h_{22}^d h_{22}^\ell \epsilon^8$,
   and with the couplings $h_{22}^d, h_{22}^\ell \sim 1$, we
    obtain the branching ratio, $BR(B_s^0\rightarrow\mu^+\mu^-)\sim 10^{-9}$.
    Current experimental limit for this BR is $4.7\times 10^{-8}$ \cite{pdg}, and
    thus this decay could be observed soon at the Tevatron as the
    luminosity accumulates.

\textbf{Vectorlike fermions, productions and decays:} Our model
 requires vectorlike quarks and leptons, both $SU(2)$ doublets, $Q_i$ and
singlets $U_i$ and $D_i$, with masses at the TeV scale. These will
 be pair produced at high energy hadron colliders via  strong
 interaction. For example, for a 1 TeV vectorlike quark, the
 production cross section at the LHC is $\sim{60}$ fb \cite {mangano}. We need
 several such vectorlike quarks for our model. So the total production
 cross section will be few hundred fb. These will decay to the light
 quarks of the same electric charge and Higgs bosons ($h$ or $s$):
 $Q\rightarrow{q h, q s}$. Thus the signal will be two high $p_T$
 jets together with the final states arising from the Higgs decay.
 For a heavy Higgs, in the golden mode ($h\rightarrow{Z Z},
 s\rightarrow{Z Z}$, this will give rise to two high $p_T$ jets plus
 four $Z$ bosons. In the case of a light $Z'$, the final state signal
 will be two high $p_T$ jets plus 8 charged leptons in the final
 state (with each lepton pair having the invariant mass of the
 $Z'$).

\section {A concrete model}

We have constructed a concrete model giving rise to the
phenomenological Lagrangian. In addition to the SM, the model has a
$U(1)_S$ local symmetry, and the $U_i(1)_{F_i}$ ($i=1,2,3$) global
symmetries. The global symmetries are slightly broken explicitly in the Higgs
potential so that there are no unwanted Goldstone bosons.  In
addition to the SM fermions, the model has a complex scalar, $S$, the
flavon fields $F_i$'s, and several vector-like quarks, both weak
doublets and weak singlets. The details of their charge assignments
under these symmetries, and how one obtains the interaction
Lagrangian can be found in \cite{lmn} .

\section{Conclusions}

We have presented a proposal in which only the top quark obtains its
mass from the Yukawa interaction with the SM Higgs boson via
dimension four operators.  All the other quarks receive their masses
from operators of dimension six or higher involving a complex scalar
Higgs $S$ whose vev is at the EW scale.  The successive hierarchy of
light quark masses is generated via the expansion parameter
$\left(\frac{S^\dagger S}{M^2}\right)\sim \epsilon^2$, where
$\epsilon \equiv \frac{v_s}{M} \sim 0.15$.  All the couplings of the
higher dimensional operators are of order one.  We are able to
generate the appropriate hierarchy of fermion masses with this small
parameter $\epsilon$.  Since $v_s$ is at the EW scale, the physics
of the new scale, $M$ is at the TeV.  Because of the new degree of
freedom at the EW scale, we have an EW singlet neutral scalar $s$,
which gives rise to interesting new physics signals which can be
tested at the LHC and at the Tevatron.  There are new scenarios for
the Higgs decays and the top quark physics. The model has a light
$Z'$ which can be produced via the Higgs decays at the LHC,  and can
give rise  invisible Higgs decays, displaced vertices for the $Z'$
decays, or  multilepton final states arise from the $Z'$ decays,
depending on the mass and lifetime of the $Z'$. We have presented a
model in which an effective interaction given in Eq. (1) can be
realized.  This requires the existence  of vectorlike quarks and
leptons, both EW doublets and singlets, at the TeV scale.  These can
be probed at the LHC.  Their decays give rise to final states with 4
$Z$'s or 4 $Z^\prime$'s and interesting new physics signals at the
LHC.

\section{Acknowledgement}
The work presented in this talk was done in collaboration with J. D.
Lykken and Z. Murdock. This research was supported in part by Grant
Numbers DOE-FG02-04ER41306 and DOE-FG02-ER46140.

\end{document}